\documentclass[aps,prl,preprint,psfig,groupedaddress]{revtex4}
\renewcommand{\vec}[1]{\mbox{\boldmath$#1$}}
\usepackage{longtable}
\usepackage{epsfig}
\def\bc{\begin{center}}
\def\ec{\end{center}}
\def\be{\begin{equation}}
\def\ee{\end{equation}}

\begin{document}

\title{Beyond the Fermi Liquid Paradigm: Hidden Fermi Liquids}

\author{J. K. Jain$^1$ and P. W. Anderson$^2$}
\affiliation{$^1$104 Davey Laboratory, Physics Department, Pennsylvania State University,
University Park, Pennsylvania 16802} 
\affiliation{$^2$Department of Physics, Princeton University, Princeton, New Jersey 08544}

\begin{abstract} An intense investigation of possible non-Fermi liquid states of matter has been inspired by two of the most intriguing phenomena discovered in the past quarter century, namely high temperature superconductivity and the fractional quantum Hall effect. Despite enormous conceptual strides, these two fields have developed largely along separate paths.  Two widely employed theories are the resonating valence bond theory for high temperature superconductivity and the composite fermion theory for the fractional quantum Hall effect. The goal of this ``perspective" article is to note that they subscribe to a common underlying paradigm: they both connect these exotic quantum liquids to certain ordinary Fermi liquids residing in unphysical Hilbert spaces.  Such a relation yields numerous nontrivial experimental consequences, exposing these theories to rigorous and definitive tests. \end{abstract}

\date{\today}

\maketitle

\section{Introduction} One may expect systems with a macroscopic number of interacting particles to be a formidable theoretical challenge, given that we do not know the exact solution of even three interacting particles.  Fortunately, several fermion liquids, for examples electrons in metals and $^3$He atoms at low temperatures, are satisfactorily described by Landau's 1956 phenomenological assumption of perturbative connectivity to a hypothetical liquid of noninteracting fermions; that is, as the interaction is slowly is turned on from zero to its full strength, the original fermions turn into weakly interacting fermionic quasiparticles, but without a drastic reorganization of the low-energy Hilbert space, or a phase transition.  This assumption enables a perturbative estimation, through the standard diagrammatic approach, of the properties of the interacting fermion liquid. Landau's Fermi liquid theory forms the basis for our understanding of a large class of fermion liquids, and has been so successful that the generic phrases ``(non) Fermi liquid" and ``quasiparticles" are often used synonymously with the more accurate expressions ``(non) Landau Fermi liquid" and ``Landau quasiparticles."   Fermi liquid theory also provides a starting point for weakly coupled superconductors and $^3$He superfluid through an instability resulting from a weak attractive interaction between the quasiparticles.   Because of their close kinship, we refer to both the normal and superconducting states of this kind as Fermi liquids in what follows.

While compelling theoretical arguments can be made for the self-consistency of the Landau Fermi liquid approach for {\em weakly} interacting fermions, its validity for real systems ultimately derives from its success in explaining experiments.  There is of course no reason to expect that every fermion liquid should conform to Landau's paradigm. The past two and half decades have witnessed an intense quest for paradigms beyond the familiar Landau Fermi liquids, especially in two dimensions, fueled by the unexpected  discoveries of two exotic phenomena, the fractional quantum Hall effect\cite{Tsui} (FQHE, 1982) and high temperature superconductivity\cite{Bednorz} (HTSC, 1986).  It has been recognized since the very outset that interactions in these systems are strong and play a nonperturbative role\cite{Laughlin,PWA87}.  Strong correlations in HTSC cuprates originate from a large onsite repulsion between two fermions (holes in the Hubbard insulator) occupying the same site, which effectively translates into an elimination of double occupancies.  In FQHE, when electrons are restricted to the lowest Landau level, as appropriate at very high magnetic fields, the only energy left in the problem is their Coulomb repulsion, which cannot be neglected and has nonperturbative consequences.  There is no small parameter in either of these problems, and it is useful to note that neither can be understood as an instability of a Fermi-liquid ``normal state."  The high T$_c$ superconductors heat, depending on parameters, into a pseudogap phase or a strange metal, neither of which exhibits Landau Fermi liquid behavior.  For the FQHE state, switching off interactions does not produce a unique Fermi-sea-like state that could serve as the normal state, but rather an exponentially divergent number of degenerate ground states; the Coulomb interaction mixes them in some complex manner to produce the FQHE.  The absence of a Fermi-liquid normal state lies at the heart of the paradigmatic crisis marked by these two phenomena. In particular, the standard perturbative approach that works well for the Landau Fermi liquids is utterly unproductive for these systems.

Much progress has been made in our understanding of both HTSC and the FQHE states during the last two decades, and a fervent search for non-Fermi liquid states in other contexts is ongoing.  At this stage, it appears meaningful to ask whether each non-Fermi liquid is to be treated individually, which is in principle possible but not particularly satisfying, or whether certain classes of non-Fermi liquids subscribe to a common new paradigm.  At first blush, the analogy between the HTSC and the FQHE states does not seem to extend beyond their non-Fermi-liquid character.  We suggest here that they are both examples of ``hidden Fermi liquids" (HFLs), a phrase introduced by one of us \cite{PWA07} to denote non-Fermi liquid states that are related to ordinary Fermi liquids residing in unphysical Hilbert spaces.  We believe that this notion has the potential for applicability to a larger class of non-Fermi liquids.  The aim of this perspective article is to illustrate in what sense the resonating-valence-bond (RVB) theory of HTSC and the composite fermion (CF) theory of FQHE describe hidden Fermi liquids, and how the connection to Fermi liquids leads to testable experimental consequences.  Fermion Jastrow wave functions, used previously for the $^3$He liquid, also relate the interacting state to a noninteracting Hartree-Fock state, but in that case the augmentation by the Jastrow factor only causes renormalizations of the Fermi liquid parameters, not a qualitative change in the nature of state.

\section{Hidden Fermi Liquid}  We begin with a brief review of the resonating valence bond and the composite fermion theories in a way that brings out their common HFL nature. This is most evident from their explicit wave functions.  Anderson's wave function for the various liquid phases of HTSC cuprates is given by \cite{PWA87,PWA07,Zhang88,Paramekanti01,PWA04,PWA06}
\be
\Psi_{\rm HTSC} = {\rm P}_{\rm GW} \Phi_{\rm HF}(\vec{r}_1, \cdots , \vec{r}_N)\;,
\ee
where $\Phi_{\rm HF}$ is a Hartree-Fock wave function for either an ordinary Fermi liquid or a BCS superconductor, and the Gutzwiller projection operator ${\rm P}_{\rm GW}$ eliminates double occupancies.  $\Psi$ can be  considered a variational wave function for the $t$-$J$ Hamiltonian acting in the space without double occupancy, a model believed to capture the relevant physics of the HTSC cuprates. This Hamiltonian can be derived  from the Hubbard model with a large on-site $U$ by making the Rice canonical transformation $\exp(iS)H\exp(-iS)$ to eliminate matrix elements connecting to states containing double occupancies \cite{Gros}.  The wave function $\Psi_{\rm HTSC}$ was originally motivated\cite{PWA87} by the RVB ideas of Fazekas and Anderson \cite{Fazekas} on a spin liquid state for frustrated antiferromagnets, and the fact that the HTSC materials are doped antiferromagnets.  

The CF wave function for the fractional quantum Hall state at filling factor $\nu$ has the form \cite{JKJ89} 
\be
\Psi_{\rm FQHE} = {\rm P}_{\rm LLL} J\Phi_{\rm HF}(\vec{r}_1, \cdots , \vec{r}_N)
\ee
where $\Phi_{\rm HF}$ is the Hartree-Fock wave function of noninteracting electrons at an effective filling factor $\nu^*$, given by $\nu=\nu^*/(2p\nu^*+1)$. The Bijl-Jastrow factor $J$ is defined as $J=\prod_{j<k}(z_j-z_k)^{2p}$, with $z_j=x_j-iy_j$, where $(x_j,y_j)$ are the coordinates of the $j$th electron in the two-dimensional plane.  The lowest Landau level projection operator ${\rm P}_{\rm LLL}$ eliminates terms with amplitude in higher Landau levels, as appropriate for very high magnetic fields.   Unification of the fractional and the integral quantum Hall effects originally served as the inspiration for the  trial wave function of this form. 

The wave function ${\rm P}_{\rm LLL}J\Phi_{\rm HF}$ describes weakly interacting composite fermions at an effective magnetic field.  The Bijl-Jastrow factor $J$ binds to each electron $2p$ quantized vortices, and the bound entity consisting of an electron and $2p$ vortices is interpreted as a particle called composite fermion.  Composite fermions are weakly interacting because the only role of interactions is to produce composite fermions through the Bijl-Jastrow factor -- the remaining factor $\Phi_{\rm HF}$ is a wave function of weakly interacting fermions.   Furthermore, the bound vortices generate Berry phases which partly cancel the Aharonov-Bohm phases produced by the external magnetic field, and composite fermions experience  an effective magnetic field $B^*=B-2p\rho\phi_0$ ($B$ is the external magnetic field, $\rho$ is the electron density, and $\phi_0=hc/e$ is the `flux quantum'), which corresponds to the filling factor $\nu^*$ defined below Eq. 2. The HFL of electrons in the FQHE regime thus reveals itself as a regular Fermi liquid of composite fermions.

A remarkable aspect of these wave functions, which is the central message of this article, is that they both establish a definite mapping of the strongly correlated non-Fermi liquid, $\Psi_{\rm HTSC}$ or $\Psi_{\rm FQHE}$, to an ordinary Fermi liquid $\Phi_{\rm HF}$ inhabiting an unphysical Hilbert space.  In HTSC the unphysical state $\Phi_{\rm HF}$ allows double occupancy, whereas in FQHE $\Phi_{\rm HF}$ is the wave function of weakly interacting fermions at an effective magnetic field.  It is stressed that the mapping between the physical state $\Psi$ and the unphysical Fermi liquid $\Phi$ applies not only for the ground state but for all low energy states, as would be necessary in order for the mapping to make thermodynamic sense; that is, all low-energy eigenfunctions of the physical system are images of the corresponding eigenfunctions in the unphysical space, with the same ordering. The excitations of an ordinary Fermi liquid ground state are constructed by application of fermion creation or annihilation operators to produce quasiparticles or quasiholes.  For our exotic liquids, the {\em physical} excitations are related to Landau quasiparticles in the same way as the ground state $\Psi$ is to $\Phi$:  the quasiparticle-like excitations are given by ${\rm P}_{\rm GW}c^\dagger \Phi_{\rm HF}$ or ${\rm P}_{\rm LLL}Jc^\dagger \Phi_{\rm HF}$, and quasihole-like excitations by analogous expressions obtained by replacing the creation operator $c^{\dagger}$ by the annihilation operator $c$ in the preceding expressions.  It is important to note, however, that the physical quasiparticles are to be distinguished from, and are not perturbatively related to, the Landau quasiparticles $c^\dagger {\rm P}_{\rm GW} \Phi_{\rm HF}$ or $c^\dagger{\rm P}_{\rm LLL}J \Phi_{\rm HF}$.

The mapping from $\Phi$ to $\Psi$ has two essential components. First, it projects out a short range part of the interaction.  The projection ${\rm P}_{\rm GW}$ in Eq. 1 eliminates high energy states residing in the upper Hubbard band, as appropriate for the Hubbard model with a large on-site Mott-Hubbard $U$.  In other words, the wave function of Eq. 1 minimizes the interaction $V_0=U\sum_{j<k}\delta^{(2)}(\vec{r}_j-\vec{r}_k)$, which penalizes electron coincidences.  The situation is more complicated but  conceptually analogous for $\Psi_{\rm FQHE}$, for which there are effectively two projections, $J$ and ${\rm P}_{\rm LLL}$; $J$ is not commonly thought of as a projection operator, but it is actually more fundamental than the lowest Landau level projection ${\rm P}_{\rm LLL}$.  To see that, let us first neglect  ${\rm P}_{\rm LLL}$ in Eq. 2.   The Bijl-Jastrow factor $J$ in the wave function $J\Phi_{HF}$ explicitly serves to project out high {\em interaction} energy configurations by restricting the Hilbert space to wave functions for which the probability of two electrons approaching one another vanishes as $r^{4p+2}$, as opposed to the usual $r^2$ dictated by the Pauli principle.  This is analogous to restricting the HTSC Fock space to the lower Hubbard band.  Formally, the wave function $J\Phi_{\rm HF}$ minimizes a short range interaction of the type\cite{TK} $V_0'=\sum_{m=1}^{p} U_m\sum_{j<k}\nabla^{2(2m-1)}\delta^{(2)}(\vec{r}_j-\vec{r}_k)$, the expectation value of which for $J\Phi_{\rm HF}$ is identically zero; $V_0'$ is the simplest generalization of the ``contact" interaction for fully polarized electrons (for which $V_0$ is invisible due to the Pauli principle).  While it minimizes the {\em interaction} energy $V_0'$, $J\Phi$ is not restricted to the lowest Landau level, as would be desirable for very high magnetic fields.  Detailed calculations show, however, that $J\Phi$ is predominantly in the lowest Landau level, and the explicit lowest Landau level projection is a final adjustment to the wave function to fit the real Hamiltonian.  This step  only causes perturbative changes, but no phase transition; a convincing case can be made that ${\rm P}_{\rm LLL}J\Phi_{\rm HF}$ is perturbatively connected to $J\Phi_{\rm HF}$, indicating that the lowest Landau level projection is important if one is interested in accurate energetics, but the essential physics of the FQHE is captured by $J\Phi_{\rm HF}$.  

The minimization of the short range part of the interaction is only half of the story, for it does not, by itself, impose any constraints on the form of $\Phi$.  A critical element of the HFL ansatz is that the physical state of the {\em full} Hamiltonian, which includes the longer range part of the interaction as well, is obtained by identifying $\Phi$ as a state of weakly interacting fermions.  This nontrivial postulate defies rigorous theoretical derivation, and can be justified, much like the Fermi liquid theory, only by comparison to experiment (and, in case of the FQHE, to exact results available for small systems).

\section{Experimental consequences of the HFL ansatz} 
If nothing else, the HFL ansatz has the virtue of presenting a precisely defined premise, the qualitative and quantitative consequences of which can, in principle, be deduced. The wave function for weakly interacting fermions, $\Phi_{\rm HF}$, can assume different forms, such as a Fermi liquid, a weakly coupled BCS superconductor, or an integral-quantum-Hall state, which produces a rich variety of strongly correlated states with a plethora of experimental consequences.  
We discuss here some of them to bring out the similarities between the HTSC and the FQHE physics, while emphasizing how the Landau picture of perturbative continuity breaks down. 

It bears mentioning that even though the RVB theory of high temperature superconductivity has inspired an enormous amount of theoretical and experimental activity, it remains controversial and not yet widely accepted by the research community.
We believe, however, that the accumulation of evidence discussed below points to its  essential correctness.

\subsubsection{Strange metals}

We begin by asking what state is produced when $\Phi_{\rm HF}$ is taken as the Hartree-Fock wave function of an ordinary, uncorrelated Fermi sea. 
The resulting state:
\begin{eqnarray}
\Psi_{\rm strage-metal}&=&{\rm P}_{\rm GW} \Phi_{\rm Fermi-sea}(\{\vec{r}_j\}) \nonumber\\
\Phi_{\rm Fermi-sea}&=& \prod_{k<k_F}c_{\vec{k}\uparrow}^{\dagger} c_{-\vec{k}\downarrow}^{\dagger} |0\rangle\
\end{eqnarray}
has been proposed to describe the ``strange metal" phase \cite{PWA06}, which is the unconventional normal state of the high T$_c$ cuprates at optimal doping.
It has been argued \cite{PWA07,PWA06} that the physical excitations ${\rm P}_{\rm GW}c^\dagger \Phi_{\rm HF}$ have a vanishing overlap with the Landau quasiparticles $c^\dagger {\rm P}_{\rm GW}\Phi_{\rm HF}$, i.e., the wave function renormalization factor $Z$ vanishes upon Gutzwiller projection, implying a non-Fermi liquid state.
The same theory makes a detailed prediction for the spectral function in the strange metal phase\cite{PWA06}; the theoretical energy distribution curves differed from the earlier angle resolved photoemission spectroscopy (ARPES) data, but are in excellent quantitative agreement with the recent, and more accurate laser-ARPES data of Ref. \cite{Dessau,Dessau2} with a single fitting parameter \cite{Casey}.

For FQHE, the state derived from the zero field Fermi sea, 
\be
\Psi_{1/2} = {\rm P}_{\rm LLL} J\Phi_{\rm Fermi \;sea}, 
\ee
describes the compressible state at $\nu=1/2$ as the composite-fermion Fermi sea \cite{HLR}.  (The effective magnetic field $B^*$ vanishes at $\nu=1/2$.)  This is analogous to the strange metal phase of HTSC. The absence of a gap in this state resolves the long-standing mystery of why no fractional plateau is seen at filling factor 1/2.  The composite-fermion Fermi sea description of the 1/2 state has been confirmed in numerous experiments.   The magnetic field experienced by the current carrying quasiparticles has been measured in several experiments \cite{Du93,Kang93,Willett93,Goldman94,Smet96,CF_review} at filling factors slightly away from $\nu=1/2$, where $B^*$ is nonzero but small; these experiments determine the radius of the cyclotron orbit by geometric means and find that it corresponds to $B^*$ rather than $B$.  Because the effective magnetic field is the fundamental defining property of composite fermions, this is a direct observation of composite fermions. The mass, magnetic moment, spin, Fermi wave vector, cyclotron resonances, Shubnikov-de Haas oscillations, and various other excitations of composite fermions have also been measured in various other experiments \cite{CF_review}.

\subsubsection{Paired state and the pseudogap phase}

For the superconducting and the pseudogap phases of HTSC, it is natural to take $\Phi$ as the unconstrained Hartree-Fock BCS wave function of a d-wave superconductor:
\begin{eqnarray}
\Psi_{\rm HTSC}&=&{\rm P}_{\rm GW} \Phi_{\rm BCS}(\{\vec{r}_j\}) \nonumber\\
\Phi_{\rm BCS}&=& \prod_{\vec{k}}\left(u_{\vec{k}}+v_{\vec{k}}c_{\vec{k}\uparrow}^{\dagger} c_{-\vec{k}\downarrow}^{\dagger}\right) |0\rangle\;.
\end{eqnarray}
This wave function is a linear superposition of terms containing singlet pairs without double occupancies -- hence the name RVB.  The consequences of this wave function have been obtained by a number of methods, including a reliable variational quantum Monte Carlo technique \cite{PWA04,Zhang88,Paramekanti01,Anderson01}.  Strong on-site repulsion naturally favors d-wave pairing, one of the otherwise puzziling aspects of the HTSC superconductors. The calculated d-wave off-diagonal long-range order parameter exhibits a dome shaped dependence on the doping $x$, consistent with phase diagram of high T$_c$ superconductivity.  A qualitative outcome of the Gutzwiller projection is that, unlike in the ``unprojected" BCS theory, the superconducting order parameter is not proportional to the d-wave gap parameter $\Delta$.  Strikingly, the variationally determined $\Delta$ remains nozero and very large (or order $J$) even as the doping vanishes, indicating a large amplitude for the gap but a loss of phase coherence.  The theoretically determined $\Delta$ falls linearly with doping ($x$) from a large value of order $J$ to zero near $x\approx 0.3$.  The simple HFL ansatz of Eq. 1 thus produces a unified description of the phase diagram of the HTSC materials as a function of doping and temperature.  The theory also provides a natural explanation for why well-defined Landau quasiparticles appear in the superconducting state:  the opening of a gap severely restricts the quasiparticle continuum producing a nonzero $Z$\cite{PWA06}.   Detailed calculations have enabled a determination of the doping dependence of the {\em nodal} quasiparticle weight $Z$ and the renormalization of its  Fermi velocity, and show a surprisingly good agreement with experiment.  Finally, because the relevant energy scale $J$ is quite large, this approach naturally produces high values of T$_c$.

It is worth noting that in the superconducting phase the Gutzwiller projected BCS wave function has the same spontaneously broken symmetry as the unprojected BCS wave function, indicating the possibility that the two might be adiabatically connected (although the issue can be a subtle one).   Nonetheless, the Gutzwiller projection results in striking renormalizations of parameters of the superconducting state, a qualitatively different behavior for the superfluid density, and in some regions of the phase diagram it causes a nonperturbative change by destroying superconductivity and producing the pseudogap phase that has a ``gap" but no off diagonal long range order.

In contrast to the CF Fermi sea at half filled lowest Landau level, a FQHE state is observed at the half filled second Landau level, i.e. at filling factor 5/2\cite{Willett}.  It is believed to be described by the so-called Pfaffian wave function \cite{Moore}
\be
\Psi_{\rm 5/2-Pfaffian}=\prod_{j<k}(z_j-z_k)^2 {\rm Pf} \left( {1 \over z_j-z_k}\right )\;.
\ee
where the Pfaffian is defined as ${\rm Pf} ( {1 \over z_j-z_k} )\equiv {\cal A} 
( \frac{1}{z_1-z_2}\cdot \frac{1}{z_3-z_4}\cdots \frac{1}{z_{N-1}-z_N} )$, 
${\cal A}$ being the antisymmitrization operator. Because ${\rm Pf} ( {1 \over z_j-z_k} )$ has the form of the BCS wave function with p-wave pairing, the wave function $ \Psi_{\rm 5/2-Pfaffian}$ represents a p-wave paired state of composite fermions.  A priori, a more natural wave function for paired composite fermions is
\begin{eqnarray}
\Psi_{\rm 5/2-BCS}&=&{\rm P}_{\rm LLL}\prod_{j<k}(z_j-z_k)^2 \Phi_{\rm BCS}
\nonumber \\
\Phi_{\rm BCS}&=&{\rm Pf} [g(\vec{r}_i-\vec{r}_j) ],
 \\
g(\vec{r}_i-\vec{r}_j)&=&\sum_{\vec{k}}g_k\phi_{\vec{k}}(\vec{r}_i)\phi_{-\vec{k}}(\vec{r}_j),\nonumber
\end{eqnarray}
where the variational parameters $g_k$ are to be determined by energy minimization.  A convincing case has been made \cite{Moller} that the Pfaffian wave function is a special case of $\Psi_{\rm 5/2-BCS}$; the latter reduces to the Pfaffian wave function for an appropriate choice of $g_k$, and to the CF Fermi sea in another limit. 

For both FQHE and HTSC, the same strong repulsive interaction that produces the exotic non-Fermi-liquid behavior can also lead to pairing without the need for an attractive interaction between electrons.  Essentially, the non-negotiability of the elimination of double occupancies in Eq. 1 and the form of the Bijl-Jastrow factor in Eq. 2 causes an {\em over}screening of the Coulomb interaction for appropriate parameters to produce an attractive interaction between the physical quasiparticles.  In HTSC materials pairing originates because the $J$ term in the $t$-$J$ model implies an attraction in the d-wave channel.  In FQHE, the interaction between composite fermions is attractive at filling $\nu=5/2$\cite{Scarola}, where a Cooper pairing of composite fermions destabilizes the composite-fermion Fermi sea, opening a gap and producing FQHE.  

\subsubsection{Fractional quantum Hall effect}

Another consequence of the HFL structure is that the mysterious phenomenon of FQHE lends itself to an explanation as the integral quantum Hall effect of composite fermions.  The integral fillings $\nu^*=n$ of composite fermions map into $\nu=n/(2pn+1)$, which are precisely the prominently observed sequences of odd-denominator fractions.   Their wave functions are related to the integral quantum Hall effect (IQHE) wave functions $\Phi^{\rm IQHE}_n$ as \cite{JKJ89}
\be
\Psi^{\rm FQHE}_{n/(2pn+1)} = {\rm P}_{\rm LLL} J\Phi^{\rm IQHE}_{n},
\ee
which have been demonstrated, by comparison to exact results on small systems,  to be extremely accurate, for both ground and excited states.  
Certain delicate FQHE states, such as 4/11, are observed only at very low temperatures and for the highest quality samples; these are explained as the {\em fractional} quantum Hall effect of composite fermions \cite{411,Chang}. 

The physical excitations of the FQHE state, ${\rm P}_{\rm LLL}Jc^\dagger \Phi_{\rm HF}$, are excited composite fermions, which have zero overlap with the Landau quasiparticles $c^\dagger  {\rm P}_{\rm LLL}J\Phi_{\rm HF}$.  The physical excitations are so different from electrons that the electron spectral function is not a useful concept for the FQHE (although it can be evaluated).  In fact, one of the remarkable properties is that when measured relative to the background FQHE state, the excited composite fermions have fractional charge excess associated with them, and are theoretically believed to obey fractional braiding statistics \cite{Laughlin,CF_review,Halperin84}.

It is straightforward to see that composite fermions are not perturbative evolutions of electrons but topologically distinct entities.  Bound states are always nonperturbative objects, and in the case of composite fermions, vortices are either bound to electrons or not -- the notion of binding them continuously is meaningless.  In a practical sense, the formation of composite fermions manifests itself directly through the qualitative feature that the dynamics of composite fermions is governed by the reduced effective magnetic field $B^*$ rather than the external magnetic field $B$.  The appearance of FQHE at the principal sequences $n/(2pn+ 1)$, terminating into a CF Fermi sea at $1/2p$, provides an experimental confirmation of the effective magnetic field.  The Fermi sea and the IQHE of composite fermions cannot be perturbatively obtained from an ordinary Fermi sea or IQHE of electrons for the simple reason that, in the physical space of the lowest-Landau level, no Fermi sea or IQHE exists for noninteracting electrons.

\section{Concluding remarks}
An appealing and satisfying conceptual structure has thus emerged from the study of two unrelated exotic quantum liquids.  The quantum fluid of interacting electrons in the lowest Landau level has been related to a weakly interacting Fermi liquid of composite fermions, and its various phases can be understood in terms of the IQHE, Fermi sea, and paired state of composite fermions.  The RVB theory similarly seeks to explain various phases of the HTSC materials through Gutzwiller projected Fermi sea and BCS wave functions. That an ordinary Fermi liquid may be buried underneath these phenomena is surprising and nontrivial, but provides an important starting point for exposing the true nature of these two complex strongly correlated systems, which would, arguably, be impossible to decipher directly in the physical space.  This makes one wonder if there exist other non-Fermi liquids that also conform to the HFL framework.

Before closing, we note the application of the hidden Fermi liquid concept to quantum spin liquids, which, to begin with, have no fermions.  Quantum spin liquids are charge insulators with local magnetic moments that exhibit no magnetic ordering even at absolute zero temperature.  The antiferromagnetic order may be hindered by quantum fluctuations, which is most likely to occur for spin-1/2 systems in low dimensions, or by geometric frustration.  The spin-1/2 Heisenberg model with antiferromagnetic coupling in one dimension is such a state, as shown by Bethe's exact solution.  The RVB state proposed in Ref. \cite{Fazekas} was a quantum spin liquid for the frustrated spin-1/2 Heisenberg system on a two dimensional triangular lattice, but the actual ground state for this system was later shown to possess antiferromagnetic order. For the even more frustrated spin-1/2 Heisenberg system on the kagom\'e lattice, no long range magnetic order is observed experimentally down to very low temperatures \cite{Exp1,Exp2}, suggesting the possible realization of a quantum spin liquid. A fruitful theoretical approach has been to introduce fictitious fermions, known as spinons, through the relation $\vec{S}_i=(1/2)c^{\dagger}_{i\alpha}\vec{\sigma}_{\alpha \beta} c_{i\beta}$, where $\vec{\sigma}$ represents the Pauli matrices, to transform the spin problem into a problem of interacting spinons; the spinons are also coupled to a U(1) gauge field which exists to eliminate the charge degree of freedom by imposing the constraint of a single spinon per site.  Ran {\em et al.} \cite{Ran} have made a strong case that the actual system is described by a Gutzwiller projected wave function of the form given in Eq.~3, with the appropriate Fermi sea in this case being the Dirac Fermi sea for spinons that experience $\pi$ flux through the hexagons \cite{Hastings}. 
A confirmation of this description would produce another realization of an HFL, attesting to the broader applicability of the concept.

\end{document}